\documentclass[aps,prl,preprint,superscriptaddress]{revtex4}
\usepackage{epsfig}
\begin{document}

\preprint{Accepted for publication in Physical Review C}

\title{Scaling of the giant dipole resonance widths in hot rotating nuclei from the ground state values}

\author{Srijit Bhattacharya}
\affiliation{Department of Physics, Darjeeling Government College, Darjeeling-734101, India}

\author{Deepak Pandit}
\affiliation{Variable Energy Cyclotron Centre, 1/AF-Bidhannagar, Kolkata-700064, India}

\author{S. Mukhopadhyay}
\affiliation{Variable Energy Cyclotron Centre, 1/AF-Bidhannagar, Kolkata-700064, India}

\author{Surajit Pal}
\affiliation{Variable Energy Cyclotron Centre, 1/AF-Bidhannagar, Kolkata-700064, India}

\author{S. R. Banerjee}
\email[e-mail:]{srb@veccal.ernet.in}
\affiliation{Variable Energy Cyclotron Centre, 1/AF-Bidhannagar, Kolkata-700064, India}


\date{\today}

\begin{abstract}
The systematics of the giant dipole resonance (GDR) widths in hot and rotating nuclei 
are studied in terms of temperature T, angular momentum J and mass A. The different experimental 
data in the temperature range of 1 - 2 MeV have been compared with the thermal shape fluctuation model (TSFM) 
in the liquid drop formalism using a modified approach to estimate the average values of T, J and A in the decay 
of the compound nucleus. The values of the ground state
GDR widths have been extracted from the TSFM parametrization in the liquid 
drop limit for the corrected T, J and A for a given system and compared with the 
corresponding available systematics of the experimentally 
measured ground state GDR widths for a range of nuclei from A = 45 to 194. Amazingly, the nature 
of the theoretically extracted ground state GDR widths matches remarkably well, though 1.5 times smaller,
with the experimentally measured ground state GDR widths consistently over a wide range of nuclei.

\end{abstract}
\maketitle

\section{introduction}

The collective excitations in nuclei, in particular, the giant
dipole resonance (GDR) have been studied in great detail over the
years to understand the complex quantal nuclear many body systems.
The phenomenon of GDR oscillations in nuclei, has been studied
extensively in cold as well as in hot and fast rotating nuclei.
In the case of GDR vibrations in cold nuclei, i.e. for the case of GDR
built on nuclear ground state, very well established systematics
for the resonance energy and its width as a function of nuclear mass exist.
The understanding of the mechanism for such a large width of the resonance 
is of particular importance as it gives an insight into the strong damping
mechanism of the collective dipole oscillations in nuclei. A systematic
study of the resonance widths at higher temperatures in nuclei in the
cases of GDR built on excited states gives us clues regarding the damping
mechanisms in hot nuclei and the interplay of the temperature and angular
momentum effects. In a particular nucleus,
the resonance energy remains more or less constant but its width
(or FWHM) increases as the temperature or the excitation energy 
of the nucleus increases. There had been a lot of experimental 
\cite{Hara01, Gaard01, Snov01, Kelly01, Thoe01} as well as 
theoretical \cite{Dang01, Orma01} activities to understand this 
increase of the GDR width with temperature in the past years.
The main experimental approach had been the heavy ion fusion reactions
populating the compound nucleus at different excitations and spins. 
The temperature dependence of the GDR width to some extent has been
explained in terms of adiabatic, large amplitude thermal fluctuations
of the nuclear shape -- the thermal shape fluctuation model (TSFM).
Though TSFM is successful to some extent, (in the 
temperature range
1 - 2 MeV and for low spins) it does not explain
the  results of \cite{Thoe01,Rathi01, Chak01, Rathi02}, particularly those at the lowest T (near T=1.0 MeV) for $^{120}$Sn.

In the past there have been several attempts to understand the 
global features of the temperature and spin dependence of the
measured GDR widths in a comprehensive manner, by parametrizing
the GDR widths in terms of the relevant macroscopic parameters, i.e.
temperature, angular momentum, nuclear mass etc. \cite{Chak02, Kasagi01, Brog01}. 
The most notable among them is the work of Kusnezov et al \cite{Kus01} in which
the GDR width $\Gamma(T, J, A)$ at a finite temperature (T) 
and spin (J) is parametrized  in terms of a reduced width, from a liquid drop (LD)
free energy consideration,
\begin{equation}
\Gamma_{red} = \left[\frac{\Gamma_{exp}(T, J, A)}{\Gamma(T, J=0, A)}\right]^{\frac{T+3T_0}{4T_0}} = 1 + \frac{1.8}{\left[1+e^{(1.3-\xi)/0.2}\right]}
\end{equation}
where, $\Gamma_{exp}(T, J, A)$ is the experimental width, the
reduced scaling parameter $\xi = J/A^{5/6}$ and 
\begin{equation}
\Gamma(T, J=0, A) = \left(6.45-\frac{A}{100}\right) \ln\left(1+\frac{T}{T_0}\right) + \Gamma_0(A) \label{eqn2}
\end{equation}
According to the authors,
$\Gamma_0(A)$ is usually extracted from the measured gound state GDR
and T$_0$ is taken as a reference temperature (=1 MeV). Surprisingly, they
used a value of $\Gamma_0$ = 3.8 MeV for $^{120}$Sn and $^{208}$Pb data
(after recalculating the nuclear temperatures)
which was smaller than an earlier description (5 MeV) \cite{Orma01}.
For other nuclei the authors themselves and others \cite{Rathi01, Chak01, Rathi02}
used the same parametrization with a wider range of values for $\Gamma_0$
(2.5 - 5.2 MeV), which were less than the ground state values, for describing the experimental GDR widths at different
temperatures and spins. This simple parametrization, however failed to explain the data at low temperature and
highest spins.

The important points, as they stand now, for the explanation 
of the temperature dependence of the GDR width in general 
and particularly within the framework
of TSFM in the LD regime, are 1) a
proper characterization of the nuclear temperature as shown by 
Kelly and others \cite{Kelly01} and 2) using a proper $\Gamma_0$ parameter in a
uniform way throughout the nuclear mass, temperature and angular momentum
range. Lately, there have been attempts, \cite{Chak01} to properly characterize 
the nuclear temperature in a
heavy ion fusion reaction. In this paper we have tried to estimate 
the proper nuclear temperature for the GDR $\gamma$-emission 
in heavy ion fusion - CN $\gamma$-decay experiments for
our recent measurements \cite{Srij01} as well as for other published results 
\cite{Rathi01, Chak01, Srij01, Kici01, Wiel01, Dreb01, Kici02, Brac01, Baum01, Kmie01, 
Brac02, Brac03, Brac04, Matt01, Came01, Came02, Schi01, Noor01}
in a unified treatment and compared them in the light
of TSFM calculations of Kusnezov et al with a uniform description
of the $\Gamma_0$ parameter in accordance with the measured systematics
of the ground state GDR widths over the entire nuclear mass range.

The same procedure is adopted while explaining our recently measured
GDR widths in $^{113}$Sb populated with high angular momenta 
($\leq$ 60 $\hbar$) and at temperatures $\leq$ 2 MeV \cite{Srij01} using a
$\Gamma_0$ = 3.8 MeV.

\section{Data Analysis}

\begin{figure}
\begin{center}
\includegraphics[height=8.5 cm, width=7 cm]{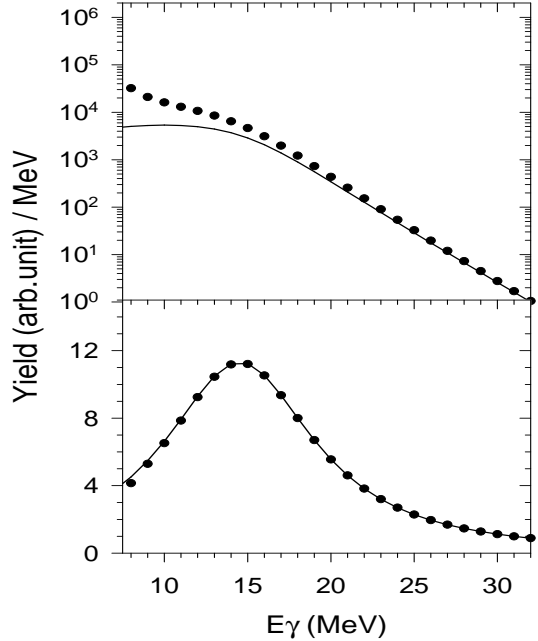}
\caption{\label{Average} The top panel shows the CASCADE predictions for
$^{113}$Sb populated at 122 MeV excitation and ~60$\hbar$ spin, with an
averaging over the full (100\%) (symbols) and part (50\%) (line) of the decay
cascade. The bottom panel shows the divided plots for the corresponding GDR 
strength distributions.}
\end{center}
\end{figure}

In heavy ion fusion - evaporation reactions, high energy $\gamma$-photons
are emitted from various stages of the decay cascade. At high excitation
energies, the compound nucleus decays through a large number of
steps and therefore, the excitation energy (E$^*$), angular momentum (J),  
mass (A) and charge (Z) should be averaged over all the decay steps.
The average values of E$^*$, J, A and Z should be different and less than 
those of the initial compound nucleus. While most authors consider an 
average temperature $\langle T\rangle$ for the corresponding measured GDR widths, the
averaging of J, A and Z has not been addressed to. Though the averaging of A and Z
does not change significantly the representation, the same is not true for J, since a small
change in $\langle J\rangle$ 
modifies the representation of the data in terms of reduced parameters
in Kusnezov's description of TSFM
\cite{Chak01}. 
Two basic approaches are generally taken in the existing literatures for the estimation of $\langle T\rangle$ for the GDR $\gamma$-emission
in a compound nucleus. In the first, $\langle T\rangle = [d ln(\rho)/ dE]^{-1}$ is
evaluated at $E^* = E_{CN} - \langle E_{rot}\rangle  - E_{GDR} -\Delta_p$, where $E_{CN}$
is the initial CN excitation energy, $\langle E_{rot}\rangle$ is the average rotational energy
computed at the mean J of the experimental J distribution and $\Delta_p$ is
the pairing energy. This procedure is incorrect since there is no averaging over E$^*$ in the CN decay chain. In the other approach, the average temperature is estimated as,
$\langle T\rangle = [(\langle E^*\rangle -\langle E_{rot}\rangle - E_{GDR} -\Delta_p)/a(\langle E^*\rangle)]^{1/2}$, 
where a($\langle E^*\rangle$) is the energy dependent level density parameter. 
In this case, though the averaging 
is done over all the decay steps, $\langle E_{rot}\rangle$ is calculated for the mean J of
the initial CN. It is also not proper to include each step in the CN decay chain for the
averaging. Instead, only that part contributing to the GDR $\gamma$-emission \cite{Wiel01}
in the decay cascade should be taken for averaging, thereby, setting a lower limit
for the excitation energy in the CN decay cascade. 
Recently, Wieland et al \cite{Wiel01} used the same procedure of averaging over
a part of the decay cascade in their analysis of highly excited $^{132}$Ce data
for a re-estimation of temperature of the nucleus emitting GDR photons.
We have followed this second
scheme of averaging to recalculate the average parameters in our work.
This lower limit in the excitation energy amounts to approximately 50\% of the total high energy $\gamma$-yield
in the CN decay chain and does not affect
the GDR $\gamma$-rays in the region E$_\gamma$ = 12 - 25 MeV. $\langle E_{rot}\rangle$ is evaluated with the average J ($\langle J \rangle$), re-estimated using the above mentioned lower limit in E$^*$. $\langle E^*\rangle$ is
calculated by averaging $E^*$ with corresponding weights over the daughter nuclei in
the CN decay cascade for the $\gamma$-emission in the GDR energy range 12 - 25 MeV,
$\langle E^*\rangle = \sum_i (E_i^* \omega_i) / \sum_i \omega_i$, where $E_i^*$ is the excitation
energy of the $i$-th nucleus in the decay cascade and $\omega_i$ is the corresponding yield in
the GDR energy region 12 - 25 MeV. The corresponding $\langle A\rangle$
and $\langle Z\rangle$ are calculated in the same way. 
Fig.\ref{Average} demonstrates the effect of averaging over the part 
of the decay cascade on the GDR strength distribution for the excited $^{113}$Sb
nucleus at an initial excitation energy and angular momentum of 122 MeV and
$\sim$ 60 $\hbar$ respectively. It is clear that this averaging leaves the GDR
strengths and high energy photon emissions unchanged, except at very low energies
without affecting the width.
The CN particle evaporation widths ($\Gamma_{ev}$)
have been incorporated in the TSFM calculation for the temperature dependence of the
GDR widths to take into consideration the effect of evaporation of particles and the corresponding
energy loss before the GDR $\gamma$-emission in the CN decay chain. The $\Gamma_{ev}$ is calculated using the modified statistical model code CASCADE \cite{Cascade} and folded with the GDR width parameter ($\Gamma_0$) in TSFM calculation. In this low temperature region (T $\leq$ 2 MeV and with $\tilde{a}$ = A/8.0),  the particle decay width is rather small ($\sim$ 0.2 MeV at T = 2.0 MeV) and its inclusion within the TSFM marginally improves the prediction. 

\begin{figure}
\begin{center}
\includegraphics[height=8.5 cm, width=7 cm]{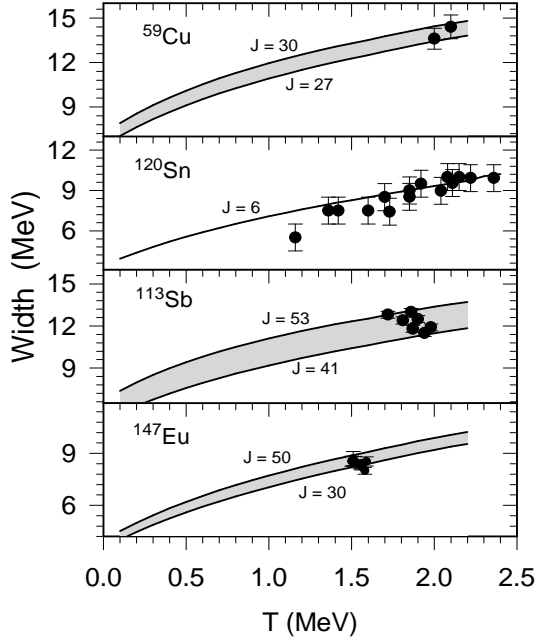}
\caption{\label{Temp}The measured GDR widths at different extracted average temperatures
are plotted for various nuclei studied. The solid lines are the predictions of the TSFM calculations
for different average angular momenta. The shaded curves show the change in the calculation
for different J values.}
\end{center}
\end{figure}

The same procedure has been used for the estimation of $\langle T\rangle$ and
$\langle J\rangle$ for nuclei in the broad mass range of 45 - 194 and temperature range
of 1 - 2 MeV (shown in tables \ref{tab:nucl}, \ref{tab:nucl1} and \ref{tab:nucl2}). All calculations have been done with a modified version of the statistical model
code CASCADE \cite{Cascade}. In almost all the nuclei, we have adopted Ignatuyk-Reisdorf level 
density prescription \cite{Igna01, Reis01}
keeping the asymptotic level density parameter $\tilde{a}$ = A/8 MeV$^{-1}$, except in the case of
$^{86}$Mo. The level density parameter for $^{86}$Mo \cite{Rathi01}
 was measured experimentally and was kept fixed at $\tilde{a}$=A/7.5 MeV$^{-1}$ as suggested by the authors.
The sensitivity of the average values has been checked by changing the
parameter $\tilde{a}$ from A/8 to A/9, resulting in a change in $\langle T\rangle \leq $ 5\%
without affecting $\langle J\rangle$ and is less than the experimental uncertainties in
measuring GDR widths. Thus it is clear that the extracted average quantities do not change
much due to uncertainties in the level density parameters.

\begin{figure}
\begin{center}
\includegraphics[height=8.5 cm, width=7 cm]{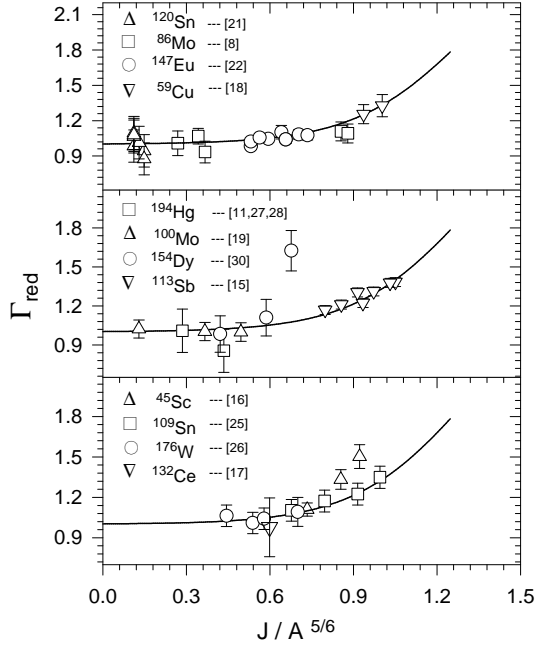}
\caption{\label{Jred}The reduced GDR widths are plotted against the reduced parameter
$\xi = J/A^{5/6}$ for different nuclei and grouped in different panels so that the data points from
different nuclei show a minimum overlap. The references for the corrosponding data points are indicated alongside the legends in the respective plots.}
\end{center}
\end{figure}

\begin{figure}
\begin{center}
\includegraphics[height=5 cm, width=7 cm]{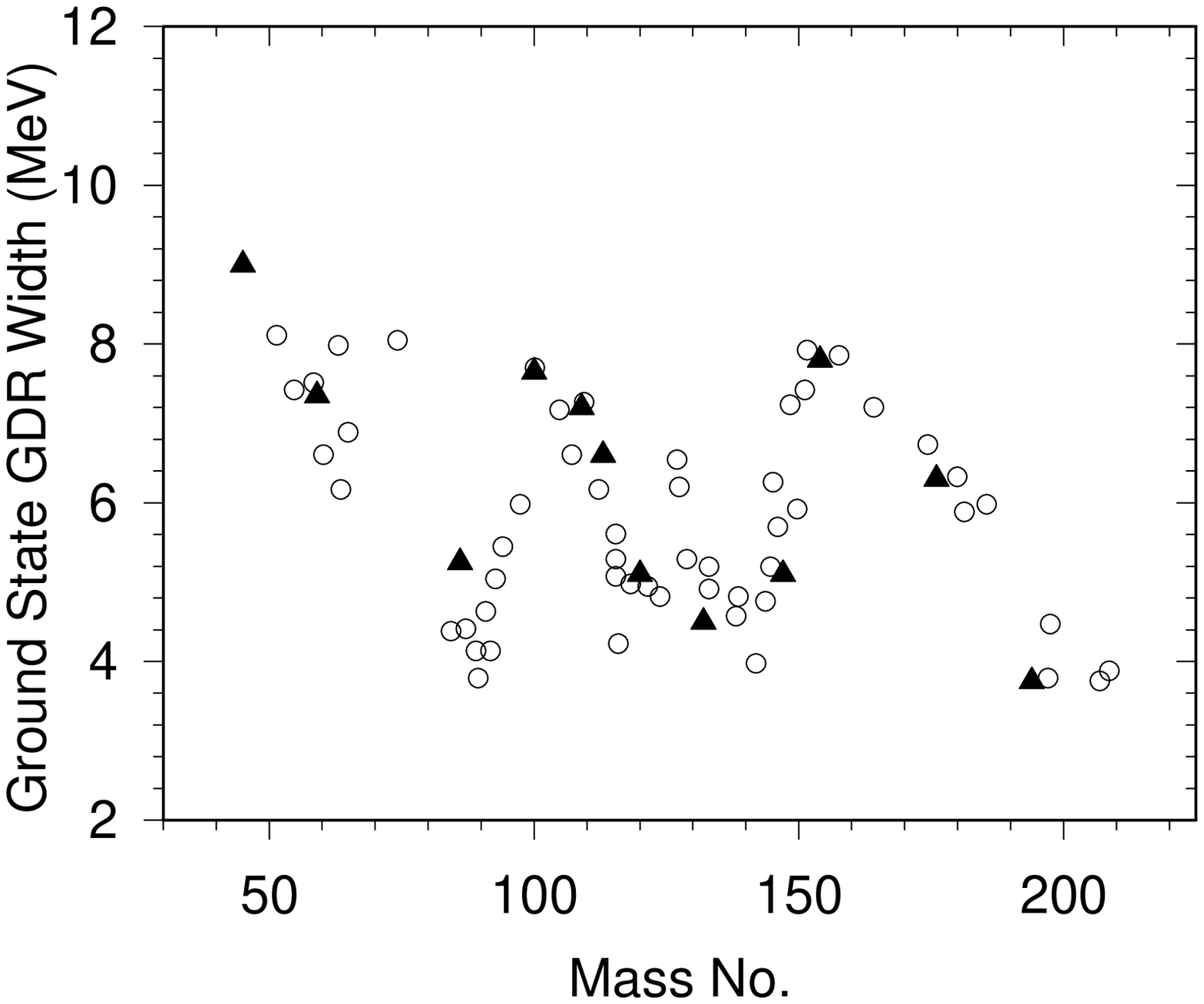}
\caption{\label{Gam}The calculated $\Gamma_0$ values are compared with the measured
ground state GDR widths (open circles) (taken from refs. \cite{Gaard01, Carl01, Diet01}) as 
a function of the nuclear mass numbers. 
The extracted $\Gamma_0$ values are also plotted (filled symbols) after multiplying by a factor of 1.5.}
\end{center}
\end{figure}

The calculations were done for all the data points following the prescription just described in
the framework of TSFM for the LD regime as suggested by Kusnezov et al. The parameter
$\Gamma_0$ was extracted for each of the nuclei studied by simultaneously fitting the
measured GDR widths at the recalculated $\langle T\rangle$ for various ranges of 
$\langle J\rangle$ (as shown in Fig.\ref{Temp}) and the reduced GDR widths $\Gamma_{red}$
at the recalculated reduced parameter $\langle J\rangle /A^{5/6}$ for different temperatures
(as shown in Fig.\ref{Jred}). Almost all the data points follow the respective calculations
quite well. The extracted $\Gamma_0$ values for different nuclei are plotted against the 
corresponding mass numbers and are shown in Fig.\ref{Gam} alongwith the measured ground
state values of the GDR widths (open circles) \cite{Gaard01, Carl01, Diet01}. 
It is clear from Fig.\ref{Gam} that the nature of
the dependence of the extracted $\Gamma_0$ values at different nuclear masses matches remarkably
well with the measured ground state GDR values. 
For comparison with the ground state GDR widths,
the extracted $\Gamma_0$ values were multiplied by a factor of 1.5. The $\Gamma_0$ 
parameter in Kusnezov's formalism is interpreted as ``the width for a spherical shape''. It is 
clear from the present representation that these $\Gamma_0$ values 
exactly follow the mass dependence of the measured
ground state GDR widths, reproducing even the effects of shell closure and nuclear shapes.
The discrepancies mentioned in Ref.\cite{Chak01} for the $^{147}$Eu and $^{154}$Dy nuclei after
re-analysis by the authors match quite well in this case. Except for one particular data point
for $^{154}$Dy \cite{Noor01} at the highest angular momentum (J=50$\hbar$)
(shown as an open circle in the middle panel of Fig.\ref{Jred})
the corresponding GDR width seems
quite large than the predicted systematics. This could be due to a rotation induced large change in
the shape of the nucleus at high excitation. The extracted $\Gamma_0$ matches remarkably well
with the ground state systematics though. The cases of $^{86}$Mo and $^{120}$Sn are particularly interesting. In the case of $^{86}$Mo the analysis reported in Ref\cite{Thoe01} uses J$_{CN}$ for calculating E$_{rot}$ and for Kusnezov's representation. Although, later, an averaging over the entire CN decay chain was done \cite{Chak01} for evaluating $\langle J \rangle$ improving the agreement with TSFM. However, the same approach could not explain the data for $^{110}$Sn. Our unified approach (averaging over a part of the decay chain for evaluating E$^*$, J, A, Z) along with a proper choice of $\Gamma_0$ explains both the data throughout. In the case of $^{120}$Sn, however, our averaging along with the corresponding $\Gamma_0$ values in accordance with the experimental ground state GDR width improves the fit overall except the points at the lowest temperatures (around T$\sim$1 MeV). It is surprising, though, the measured width at around T$\sim$ 1 MeV is smaller than that at T=0 MeV, the experimental ground state GDR width. 

The method of averaging adopted in this work, in a unified way, applied 
over a range of nuclei, is more appropriate for nuclei populated with a large
angular momentum and excitation energy. Table \ref{part} shows the extracted
average values for the two nuclei $^{86}$Mo and $^{113}$Sb populated at the 
two extremes of excitation energy and angular momentum within our data samples,
with a partial and a full average over the decay cascade. The extracted 
$\Gamma_0$ values obtained from a simultaneous description of the dependence
of experimental GDR widths in terms of $\langle T\rangle$ and 
$\langle J\rangle/A^{5/6}$ (obtained with a partial average over the decay
cascade) match quite well with the systematic dependence of experimental
ground state GDR widths as a function of nuclear mass (A).

\section{Summary and Conclusion}
In conclusion, we have studied the systematics of the GDR widths at different spins and particularly
in the temperature range of 1 - 2 MeV over a broad range of nuclear masses in the framework of the
liquid drop thermal shape fluctuation theory. The phenomenological description given by Kusnezov
et al \cite{Kus01} describes quite well all the data from various experiments done earlier even at low temperature and highest spins, provided
the temperature and the angular momentum of the decaying nucleus populated in such a heavy ion 
reaction is characterized properly using the averaging procedure discussed here and
using a $\Gamma_0$ parameter which is 1.5 times smaller than the ground state GDR width for that
particular nucleus. The extracted values of $\Gamma_0$ match exactly in shape and form
(apart from a normalization factor of 1.5) with the measured systematics of the ground state GDR widths
in spite of using a thermal fluctuation model in the liquid drop limit. The reason remains an
interesting question to investigate.

\begin{table}

\caption[Re-estimated parameters for different nuclei]{\label{tab:nucl} The re-estimated parameters using our modified approach shown along with experimental GDR widths (of A=45-113) in CN  reactions.}
		
\begin{tabular}{|c|c|c|c|c|c|c|c|c|}
\hline
CN &	E$_{beam}$ &	Ex & J$_{CN}$ &	FWHM & $\overline{J}$ &	$\overline{T}$ &	Width ($\Gamma$) & $\Gamma_0$ \\
   &  (MeV) & (MeV) & $\hbar$ & $\hbar$ & $\hbar$ & (MeV) & (MeV) & (MeV) \\
\hline
$^{45}$Sc & 72.5	& 66.6	& 18.5	& 20	& 17.5	& 1.8$^{+0.1}_{-0.15}$ & 	13.5 $\pm$ 0.5	&  \\
Ref.\cite{Kici01}	& 89.9 & 76.7&	21.4&	20&	20.4&	2.0$^{+0.15}_{-0.5}$ &	16.1 $\pm$ 0.7 & 6.2 \\
		& 109.6	&88.9	&23.5	&20	&22	&2.25$^{+0.15}_{-0.1}$	&18.1$\pm$0.9	&\\
&&&&&&&&\\
\hline
$^{59}$Cu &175	&93.2	&27.5	&20	&26.5	&2.0$^{+0.10}_{-0.05}$	&13.6$\pm$ 0.7 &	4.9 \\
Ref.\cite{Dreb01}	&215	&111.4	&31.5	&20	&30	&2.1$^{+0.05}_{-0.11}$	&14.4$\pm$0.8 & \\
&&&&&&&&\\
\hline
&&&16	&16.0	&14	&1.35$^{+0.05}_{-0.13}$	&8.8$\pm$0.5 & \\
 &100 &49 &13	&16.0	&11	&1.42$^{+0.02}_{-0.15}$	&8.5$\pm$0.8 & \\
 
$^{86}$Mo && &17	&17.0	&15	&1.32$^{+0.02}_{-0.15}$	&7.7$\pm$0.7 & 3.5 \\
\cline{2-8} 
Ref.\cite{Rathi01}&125 &66 &38	&20.0	&35	&1.23$^{+0.05}_{-0.10}$	&8.8$\pm$0.6 & \\
&&& 39	&20.0	&36	&1.24$^{+0.06}_{-0.11}$	&8.7$\pm$0.6 & \\
&&&&&&&&\\
\hline
$^{100}$Mo &49.1	&48.1	&9.3	&20	&6	&1.28$^{+0.07}_{-0.05}$	&9.79$\pm$0.62 & \\
Ref.\cite{Kici02} &63.4	&59.8	&19.5	&20	&17	&1.40$^{+0.10}_{-0.05}$	&9.90$\pm$0.63 & 5.1 \\
&72.8	&67.5	&24.0	&20	&23	&1.49$^{+0.06}_{-0.10}$	&10.06$\pm$0.64 & \\
&&&&&&&&\\
\hline

$^{109}$Sn &197	&80.2	&40	&18	&34	&1.60$^{+0.13}_{-0.05}$	&10.8$\pm$0.60 &4.8 \\
Ref.\cite{Brac04}& & &49 &16 &46  &1.40$^{+0.08}_{-0.03}$ &11.4$\pm$0.60& \\
&&&&&&&&\\
\hline
$^{110}$Sn &212	&90.1	&44	&16	&40	&1.76$^{+0.15}_{-0.02}$	&11.7$\pm$0.60 & 4.8\\
Ref.\cite{Brac04} & & &54 &14 &50  &1.57$^{+0.12}_{-0.01}$ &12.8$\pm$0.60 &\\
&&&&&&&&\\
\hline
\end{tabular}
\end{table}

\begin{table}[t]

\caption[Re-estimated parameters for different nuclei]{\label{tab:nucl1} The re-estimated parameters using our modified approach shown along with experimental GDR widths (of A=113-176) in CN  reactions.}

\begin{tabular}{|c|c|c|c|c|c|c|c|c|}
\hline
CN &	E$_{beam}$ &	Ex & J$_{CN}$ &	FWHM & $\overline{J}$ &	$\overline{T}$ &	Width ($\Gamma$) & $\Gamma_0$ \\
   &  (MeV) & (MeV) & $\hbar$ & $\hbar$ & $\hbar$ & (MeV) & (MeV) & (MeV) \\

\hline
 &145	&109	&49	&24	&41	&1.94$^{+0.06}_{-0.1}$	&11.5$\pm$0.25 & \\
           &	  &	    &53	&22	&48	&1.87$^{+0.06}_{-0.1}$	&11.8$\pm$0.25 &  \\
           &	  &	    &57	&18	&50	&1.81$^{+0.03}_{-0.1}$	&12.4$\pm$0.25 & \\
$^{113}$Sb           &    &     &60 &16 &54 &1.72$^{+0.03}_{-0.1}$  &12.8$\pm$0.25 &4.4\\ 
\cline{2-8}
Ref.\cite{Srij01}     &160 &122  &50 &24 &44 &1.98$^{+0.14}_{-0.05}$  &11.9$\pm$0.25 &\\
           &    &     &54 &20 &47 &1.90$^{+0.13}_{-0.04}$  &12.5$\pm$0.25 &\\
           &    &     &59 &18 &53 &1.86$^{+0.09}_{-0.14}$  &13.0$\pm$0.25 &\\      
&&&&&&&&\\
\hline

$^{132}$Ce &300	&100	&45	&20	&35	&1.63$^{+0.07}_{-0.05}$	&8.0$\pm$1.5 &3.0 \\
Ref.\cite{Wiel01}      &    &     &   &   &   &                       &            &    \\
\hline

   &        &       &37       &30       &34       &1.58$^{+0.01}_{-0.13}$&8.0$\pm$0.2&\\
    &  160   &73.8   &42       &20       &38      &1.56$^{+0.01}_{-0.14}$&8.4$\pm$0.2&\\    &        &       &46       &20       &41       &1.51$^{+0.04}_{-0.11}$&8.7$\pm$0.4&\\
   \cline{2-8}
$^{147}$Eu&  & &38&30&34&1.56$^{+0.02}_{-0.16}$&8.24$\pm$0.2&\\
Ref.\cite{Kmie01}&165 &77.6 &44&20&42&1.55$^{+0.02}_{-0.13}$&8.34$\pm$0.3&3.4\\
     & & &49&22&45&1.50$^{+0.05}_{-0.15}$&8.55$\pm$0.29&\\
     \cline{2-8}
     &&&39&30&36&1.59$^{+0.03}_{-0.10}$&8.53$\pm$0.26&\\
     &170  &81.4    &45&20&42&1.56$^{+0.02}_{-0.05}$&8.37$\pm$0.25&\\
     &  &    &50&26&47&1.52$^{+0.03}_{-0.04}$&8.54$\pm$0.28&\\ 
     \cline{2-8}
  \hline
 $^{154}$Dy &	&	&32	&18	&28	&1.42$^{+0.09}_{-0.07}$	&9.4$\pm$1.2 & 4.8\\
Ref.\cite{Noor01} &167 &69 &41 &26 &39  &1.33$^{+0.09}_{-0.07}$ &10.3$\pm$1.2 &\\
      & & &50& 18&45&1.25$^{+0.09}_{-0.07}$ &14.5$\pm$1.3 &\\ 
\hline
$^{176}$W&147&73.4&36&16&33&1.46$^{+0.09}_{-0.11}$ &8.9$\pm$0.6 &\\
Ref.\cite{Matt01}&&&42&15&40&1.41$^{+0.11}_{-0.13}$ &8.4$\pm$0.6 &\\
&&&47&15&43&1.39$^{+0.13}_{-0.12}$ &8.6$\pm$0.6 &4.2\\
&&&55&19&52&1.35$^{+0.08}_{-0.1}$ &8.9$\pm$0.8 &\\
\hline

\end{tabular}
\end{table}

\begin{table}[t]
\caption[Re-estimated parameters for different nuclei]{\label{tab:nucl2} The re-estimated parameters using our modified approach shown along with experimental GDR widths of A=194 in CN reaction and of A=120 in inelastic scattering reaction.}
\begin{tabular}{|c|c|c|c|c|c|c|c|c|}
\hline
CN &	E$_{beam}$ &	Ex & J$_{CN}$ &	FWHM & $\overline{J}$ &	$\overline{T}$ &	Width ($\Gamma$) & $\Gamma_0$ \\
   &  (MeV) & (MeV) & $\hbar$ & $\hbar$ & $\hbar$ & (MeV) & (MeV) & (MeV) \\
\hline

$^{194}$Hg&142&60&24&14&23&1.4$^{+0.05}_{-0.15}$ &6.5$\pm$1.0 &\\
Ref.\cite{Came01, Came02,Chak02}&&&36&8&34&1.35$^{+0.03}_{-0.10}$ &5.5$\pm$1.0 &5.1\\
&&&&&&&&\\

\hline
           &	   &34.4	&	&	&8	&1.16$^{+0.04}_{-0.11}$	&5.5$\pm$1.0 & \\
           &     &44.4 & & &8  &1.36$^{+0.04}_{-0.14}$ &7.5$\pm$1.0 & \\
           &     &54.4 & & &8  &1.60$^{+0.02}_{-0.15}$ &7.5$\pm$1.0 & \\
           &     &64.4 & & &6  &1.70$^{+0.08}_{-0.13}$ &8.5$\pm$1.0 & \\
$^{120}$Sn &200  &74.5 &10 & &7  &1.85$^{+0.11}_{-0.15}$ &8.5$\pm$1.0 & 3.4\\
Ref.\cite{Baum01}      &     &84.5 & & &6  &1.92$^{+0.11}_{-0.15}$ &9.5$\pm$1.0 & \\
					 &     &94.5 & & &6  &2.08$^{+0.14}_{-0.15}$ &10.0$\pm$1.0 & \\
					 &     &104.5 & & &6  &2.15$^{+0.15}_{-0.11}$ &10.5$\pm$1.0 & \\

&&&&&&&&\\

	   &	 &44.7 &6.36   &	&4	&1.42$^{+0.05}_{-0.10}$	&7.50$\pm$1.0 & \\
           &     &55.2 &7.68   &        &6      &1.73$^{+0.04}_{-0.12}$ &7.42$\pm$1.0 & \\
$^{120}$Sn & 160    &65.1 &9.34   &        &7      &1.85$^{+0.04}_{-0.11}$ &8.52$\pm$1.0 & \\
Ref.\cite{Rama01}& \&     &74.7 &11.05  &  3     &9      &2.04$^{+0.06}_{-0.13}$ &8.97$\pm$1.0 &  3.4 \\
           & 200   &84.5 &12.70  &        &10     &2.11$^{+0.11}_{-0.12}$ &9.55$\pm$1.0 & \\
           &     &94.9 &14.56  &        &10     &2.22$^{+0.10}_{-0.15}$ &9.92$\pm$1.0 & \\
           &     &104.8 &16.46 &        &11     &2.36$^{+0.10}_{-0.12}$ &9.92$\pm$1.0 & \\

\hline
\end{tabular}			 
\end{table}

\begin{table}
\caption{\label{part} The average values of J and T calculated with a partial (50\%) and full (100\%) 
average over the decay cascade for the two nuclei populated at about the two extremes of 
excitation energy and angular momentum.}
\begin{tabular}{|c|c|c|c|c|c|c|}
\hline
     CN   & E*   & J$_{CN}$ & $\overline{J}$ 100\% & $\overline{J}$ 50\% & $\overline{T}$ 100\% & $\overline{T}$ 50\% \\
          & (MeV)& ($\hbar$)  & ($\hbar$)  & ($\hbar$)  & (MeV)  & (MeV)     \\
\hline
          &      & 16     & 12           & 14          & 1.20         & 1.35  \\
$^{86}$Mo & 49   & 13     & 10           & 11          & 1.22         & 1.42  \\
Ref.\cite{Rathi01}&      & 17     & 13           & 15          & 1.18         & 1.32  \\ 
\hline
          &      & 50     & 40           & 44          & 1.84         & 1.98  \\
$^{113}$Sb& 122  & 54     & 44           & 47          & 1.80         & 1.90  \\
Ref.\cite{Srij01}&      & 59     & 49           & 53          & 1.78         & 1.86  \\
\hline
\end{tabular}
\end{table}



%


\end{document}